# Projectile Transverse Momentum Controls Emission in Electron Vortex Ionization Collisions

A. Plumadore and A. L. Harris*

Physics Department, Illinois State University, Normal, IL, USA 61790

**Abstract**

The realization of electron vortex beams in the past decade has led to numerous proposed applications in fields from electron microscopy to control and manipulation of individual molecules. Yet despite the many unique characteristics and promising advantages of electron vortex beams, such as transverse momentum and quantized orbital angular momentum, there remains a limited understanding of their fundamental interactions with matter at the atomic scale. Collisions between electron vortex projectiles and atomic targets can provide some insight into these interactions and we present here fully differential cross sections for ionization of excited state atomic hydrogen targets using electron vortex projectiles. We show that the projectile's transverse momentum causes the ionized electron angular distributions to be altered compared to non-vortex projectiles and that the ionized electron's ejection angle can be controlled by adjustment of the vortex opening angle, a feature unique to vortex projectiles. Additionally, an inherent uncertainty in the projectile's momentum transfer leads to a broadening of the classical binary peak, making signatures of the target electron density more readily observable. Fully differential cross sections for aligned 2p targets exhibit structures that can be used to determine the alignment.

## 1. Introduction

For many decades, fundamental discoveries about the structure of atoms and molecules have been made through the field of charged particle collisions [1]. These studies have provided an invaluable amount of information about electron charge cloud distributions and Coulomb interactions in few-body systems. Despite their long history, atomic collisions are still providing new insights, and even surprises, thanks to improved theoretical methods and advanced experimental technologies. In recent years, the COLTRIMS experimental technique has driven advancements by providing unprecedented detailed measurements [2]. Complementary to this, many theoretical models such as Exterior Complex Scaling [3], Convergent Close Coupling [4], Time Dependent Close Coupling [5], and R-Matrix with pseudostates [6] models are now considered numerically exact for some collision processes, essentially solving the 3-body problem. Looking ahead, electron vortex beams may provide the next leap forward in atomic and molecular collisions, providing a new probe of atomic structure and charged particle dynamics.

*corresponding author alharri@ilstu.edu

Electron vortex (EV) beams are matter waves with non-zero orbital angular momentum and transverse linear momentum. They have recently been experimentally realized by several groups [7–11], and may provide the opportunity for control and rotation of nanoparticles [12–15], improved resolution in electron microscopy [12,16,17], as well as the study of fundamental atomic properties, such as the magnetic moment and electronic transitions [12,13,18]. The development of EV beams was inspired by their optical counterparts, which have been widely studied [19] and are used extensively in applications such as optical tweezers [20,21], microscopy [22,23], micromanipulation [24] and astronomy [25]. EV beams, however, provide advantages their photonic counterparts cannot, such as smaller wavelengths that allow for more precise interactions with small molecules [26]. EV beams also inherently carry charge, leading to electric and magnetic effects that can be exploited to improve microscopy applications [27].

While *optical* vortex beams have a long history of study and successful application, the study and application of EV beams is still in its infancy. The proposed applications of EV beams are far-reaching and development of these applications requires a solid understanding of their interactions at the atomic scale, with atomic collision cross sections providing a vital piece of the puzzle. In addition, EV beams present a new tool for atomic and molecular collision physics itself to more intimately explore the fundamental interactions and structures of the particles in these collisions.

In [28,29], we showed that the FDCS for ionization of ground state hydrogen using EV projectiles were significantly altered compared to their non-vortex counterparts. We present here theoretical fully differential cross sections (FDCS) for ionization of atomic hydrogen from the first two excited states by EV projectiles. The present results show clear signatures of the target state structure that are not visible in the FDCS of non-vortex projectiles. We also show that FDCS for EV projectiles with carefully chosen physical characteristics can be used to identify the orientation of a spatially asymmetric target. These results are an analogue and initial test case for future studies aimed at identifying molecular structure and orientation. Through kinematical arguments, we explain the qualitative structures observed in the FDCS and show how these features can be traced to characteristics of either the target atom or vortex projectile. Atomic units are used throughout.

## 2. Theory

Electron vortex beams are experimentally generated using high energy electrons on the order of a few keV, making the first Born Approximation (FBA) sufficient for the calculation of fully differential cross sections (FDCS). For a traditional collision calculation within the FBA, the projectile is treated as a plane wave, however, for EV projectiles, a Bessel wave function is used. This Bessel wave function is a free particle solution to the Schrödinger equation in cylindrical coordinates. Unlike the traditional plane wave function, the Bessel wave function is not uniform in the direction transverse to propagation, but instead has a well-defined center. Therefore, it is possible for the target atom to be transversely offset from the Bessel wave function's center by a non-zero impact parameter. In an experiment, it is not possible to control this impact parameter for every collision event, making an average over impact parameters necessary. In this case, the FDCS in the FBA can be conveniently written in terms of the non-vortex plane wave transition matrix $T_{fi}^{PW}(\vec{q})$ [28,30]

$$\frac{d^3\sigma}{d\Omega_1 d\Omega_2 dE_2} = \mu_{pa}^2 \mu_{pi} \frac{k_f k_e}{(2\pi) k_{zi}} \int d\phi_{k_i} |T_{fi}^{PW}(\vec{q})|^2, \qquad (1)$$

where $T_{fi}^{PW}(\vec{q})$ is calculated for incident $\chi_{\vec{k}_i}$ and scattered $\chi_{\vec{k}_f}$ plane waves

$$T_{fi}^{PW} = -(2\pi)^2 <\chi_{\vec{k}_f} \chi_{\vec{k}_e} |V_i| \chi_{\vec{k}_i} \Phi>. \qquad (2)$$

The momenta of the incident projectile, scattered projectile, and ionized electron are $\vec{k}_i, \vec{k}_f, \vec{k}_e$ respectively and $\vec{q} = \vec{k}_i - \vec{k}_f$ is the momentum transfer vector. The reduced masses of the projectile and target atom and the proton and ionized electron are $\mu_{pa}$ and $\mu_{pi}$ respectively. The target hydrogen atom wave function $\Phi$ is known analytically, and the perturbation is simply the Coulomb interaction between the projectile and target atom

$$V_i = -\frac{1}{r_1} + \frac{1}{r_{12}}, \qquad (3)$$

where $r_1$ is the projectile-nuclear distance and $r_{12}$ is the projectile-target electron distance. For the kinematics considered here, the ionized electron is much slower than the projectile, and we have improved upon the model in [28] by now representing the ionized electron by a Coulomb wave in the field of the residual H$^+$ ion

$$\chi_{\vec{k}_e} = \Gamma(1-i\eta)e^{-\frac{\pi\eta}{2}}e^{i\vec{k}_e\cdot\vec{r}_2}F_1(i\eta, 1, -ik_e r_2 - i\vec{k}_e\cdot\vec{r}_2), \tag{4}$$

where $\Gamma(1-i\eta)$ is the gamma function and $\eta$ is the Sommerfeld parameter.

In the collision system used here, we define the coordinate system such that a non-vortex projectile has incident momentum along the z-axis. Following the collision, it scatters into the x-z plane with scattering angle $\theta_s$ and positive x-coordinate. The x-z plane contains the non-vortex incident, scattered, and momentum transfer vectors and is defined as the scattering plane. For a vortex projectile, the transverse momentum is non-zero, with a magnitude of

$$k_{i\perp} = k_i \sin\alpha, \tag{5}$$

where $\alpha$ is referred to as the beam's opening angle. This results in momentum transfer vector components

$$q_x = k_i \sin\alpha \cos\phi_{k_i} - k_f \sin\theta_s \tag{6}$$

$$q_y = k_i \sin\alpha \sin\phi_{k_i} \tag{7}$$

$$q_z = k_i \cos\alpha - k_f \cos\theta_s, \tag{8}$$

where $\phi_{k_i}$ is the incident momentum azimuthal angle. One of the inherent features of a vortex projectile is that the transverse momentum is not well-defined, and therefore the momentum transfer $\vec{q}$ is also uncertain. This uncertainty in the momentum transfer is accounted for by averaging the FDCS over the incident momentum azimuthal angle (see Eq. (1)) and has a significant effect on the FDCS, as shown below.

For each azimuthal angle $\phi_{k_i}$, there is a unique momentum transfer vector, which has its tail at the origin. The set of vortex momentum transfer vectors then form a cone with the point at the origin. If the heads of the vectors forming the cone are projected onto the scattering plane, they lie along a line parallel to the x-axis. In other words, for a given opening angle, the longitudinal components of the vortex momentum transfer vectors are constant. Figure 1 shows a plot of the projection of the heads of the momentum transfer vectors in the scattering plane as a function of $\phi_{ki}$ for several opening angles. For a non-vortex projectile, there is only a single, well-defined momentum transfer vector shown in Figure 1 as a solid black arrow. We refer to this as the classical momentum transfer. If $\phi_{k_i} = 0$ or $\pi$, the vortex momentum transfer vector lies in the scattering plane, while for all other values of $\phi_{k_i}$, the momentum transfer vector points outside the scattering plane. As the opening angle increases, the spread of the vortex momentum transfer vectors increases, resulting in larger uncertainty in the momentum transfer vector. Additionally, for an opening angle greater than the scattering angle, some of the vortex momentum transfer vectors have a transverse component in the same direction as the scattered projectile (positive x-component). We show below that this has a significant effect on the ejected electron angular distributions. These features of momentum transfer uncertainty hold for all excited states of the target atom.

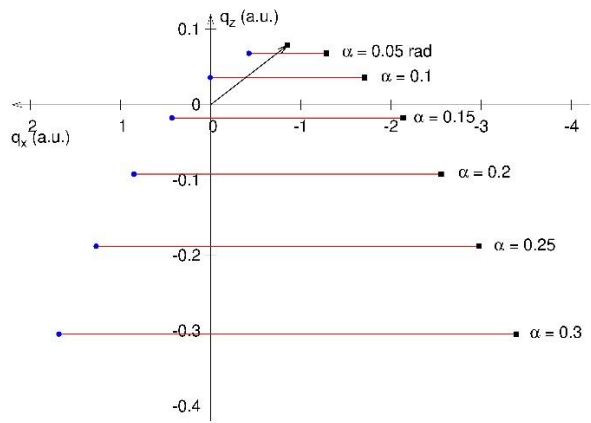

Figure 1 Momentum transfer vector projections in the scattering plane for an electron vortex projectile for all possible $\phi_{ki}$. Each opening angle $\alpha$ has a set of momentum transfer vectors that lie on a cone. The heads of these vectors appear as a line (red) when projected onto the scattering plane. For an incident plane wave, there is only one momentum transfer vector (black arrow). The blue circles at the minimum $q_x$ values occur for $\phi_{ki} = 0$ and the black squares at the maximum $q_x$ values occur for $\phi_{ki} = \pi$. Results are shown for an incident projectile energy of 1 keV, scattering angle 100 mrad, ionized electron energy 5 eV, and $n = 2$ hydrogen target.

## 3. Results

For FDCS with EV projectiles averaged over impact parameter, the only physical parameter distinguishing the EV projectile from that of a plane wave is the opening angle $\alpha$, which determines the projectile's transverse momentum. If $\alpha = 0$, the plane wave projectile is recovered and the incident projectile has only longitudinal momentum. For $\alpha \neq 0$, the incident vortex projectile has both longitudinal and transverse momentum $k_{i\perp}$, leading to the uncertainty in momentum transfer discussed above. This uncertainty has the effect of broadening the main peak in the FDCS as observed in Figure 2, which shows the FDCS for ionization of hydrogen from the ground and first two excited states as a function of ejected electron angle and opening angle. The kinematics were chosen such that future experiments may be possible (high incident energy) and the first Born approximation is applicable (asymmetric outgoing electron energies and small perturbation parameter). For these FDCS, a 1 keV incident electron scatters from the target at a fixed scattering angle of 100 mrad (5.73°); the ionized electron has an energy of 5 eV. We note that the FDCS for other kinematical parameters exhibit the same qualitative behaviors observed here, and we begin with a discussion of some qualitative features present for all target states.

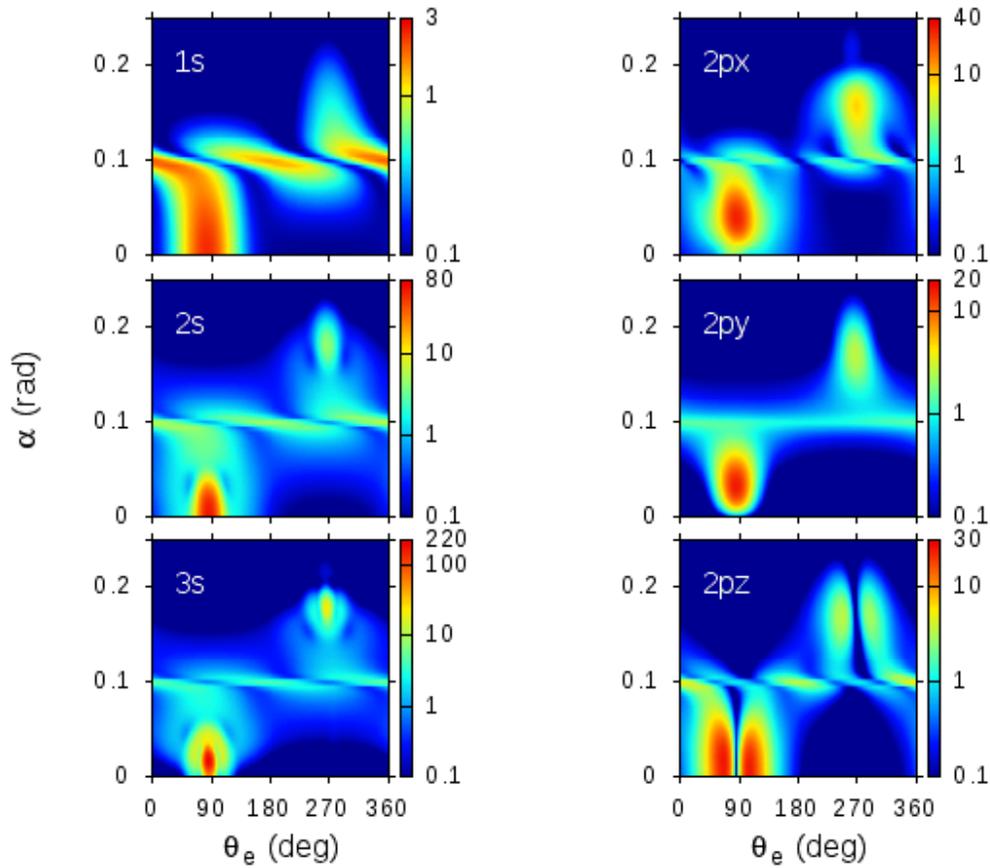

Figure 2 Fully Differential Cross Sections for ionization of H by EV projectile as a function of EV opening angle $\alpha$ (vertical axis) and ionized electron angle $\theta_e$ (horizontal axis). The color bar represents the magnitude of the FDCS. The incident projectile energy is 1 keV, scattering angle is 100 mrad, and ionized electron energy is 5 eV. The target state is shown in the figure.

For small values of $\alpha$ ($\lesssim 0.05$ rad), the traditional binary peak due to a direct collision between the projectile and target electron is observed along the classical momentum transfer direction ($\theta_e = 81.8°$ for $n = 1$, $\theta_e = 84.7°$ for $n = 2$, and $\theta_e = 85.3°$ for $n = 3$). However, no recoil peak along the direction opposite the classical momentum transfer direction is present due to the kinematics. As is well-understood in plane wave collisions, if the momentum transfer $\vec{q}$ and initial target electron momentum are well-defined, then the binary peak would be sharp. However, the momentum distribution of the initial target electron results in a broad binary peak centered about the classical momentum transfer direction. The same broad binary peak is also observed for vortex collisions, however as the vortex opening angle $\alpha$ increases, the binary peak broadens even more due to the uncertainty in the momentum transfer. The location of the binary peak also shifts to smaller angles as opening angle increases.

For $\alpha > 0.1$ rad, the dominant peak is observed at the classical recoil peak direction, opposite to the classical momentum transfer direction. This shift is a result of one momentum transfer direction being dominant in the average over azimuthal angles. Because each non-vortex FDCS used in the average depends inversely on powers of the momentum transfer magnitude, FDCS from azimuthal angles resulting in smaller momentum transfer magnitude will dominate the average. Figure 3 shows the magnitude of the momentum transfer as a function of opening angle for all projectile momentum azimuthal angles. From this, it is clear that for a given opening angle, the smallest value of momentum transfer magnitude occurs for $\phi_{ki} = 0$ and the largest for $\phi_{ki} = \pi$. This indicates that the non-vortex FDCS for $\phi_{ki} \approx 0$ dominates the average, with the contributions of all other FDCS diminishing rapidly as $\phi_{ki}$ moves away from 0. Because $\phi_{ki} = 0$ is dominant in the average of the FDCS, the direction of the momentum transfer for this particular azimuthal angle is the primary influence of the peak location in the FDCS. It can be seen from Fig. 1 that the momentum transfer direction for $\phi_{k_i} = 0$ (blue circles) shifts to a more forward direction as $\alpha$ increases toward 0.1 rad, is exactly forward ($q_x = 0$) at $\alpha = 0.1$ rad, and then is oriented backward for $\alpha > 0.1$ rad, which nicely correlates with the FDCS peak location.

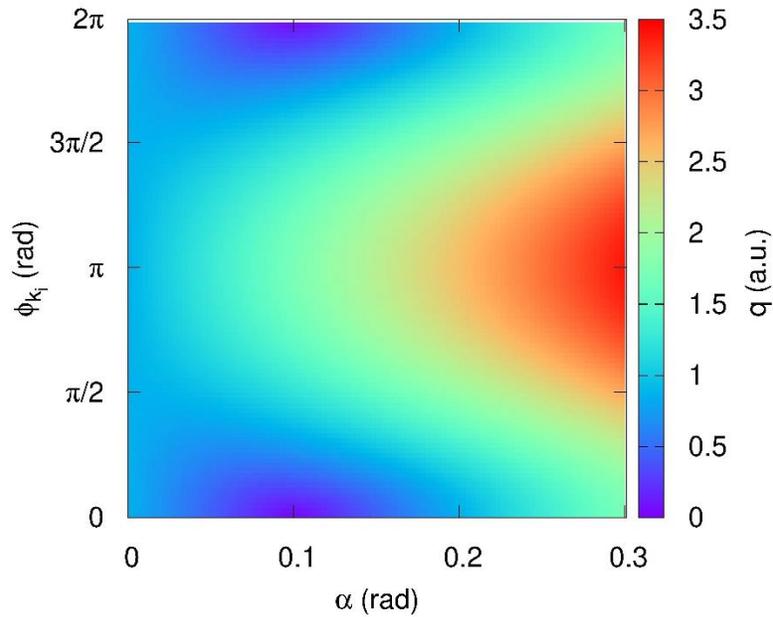

Figure 3 Magnitude of momentum transfer as a function of projectile opening angle $\alpha$ and incident momentum azimuthal angle $\phi_{k_i}$. The incident projectile energy is 1 keV, scattering angle 100 mrad, ionized electron energy 5 eV, and $n = 2$ hydrogen target.

A plot of the momentum transfer angle for $\phi_{ki} = 0$ compared to the peak location of the FDCS is shown in Fig. 4 for the ground state, where it is readily observable that the FDCS peak is almost exactly predicted by the momentum transfer direction for $\phi_{k_i} = 0$. Similar results are found for the all but the 2pz excited state, which exhibits a double peak structure centered around the classical momentum transfer direction for $\phi_{k_i} = 0$. This explains the transition from a dominant peak along the classical binary peak direction to along the classical recoil peak direction, as seen in Figure 2. In fact, this "recoil" peak is actually the binary peak caused by a momentum transfer vector resulting from the projectile being deflected toward the beam direction (-x direction) rather than away from it (+x direction). This is a feature only possible with EV projectiles, as a non-vortex projectile is always deflected toward the +x direction for FDCS with fixed scattering angle and the geometry defined here. The shift in location of the binary peak provides a possible means to control ionized electron emission angle for fixed energies and scattering angle, which could have potential applications in electron microscopy or collisions with delicate targets. In these cases, low energy electrons may cause additional noise in a signal or possible damage to a target or sample. If the secondary electron could be primarily emitted into a region not of interest, the signal to noise ratio may be enhanced or the lifetime or survivability of the sample improved.

The FDCS for $\alpha = 0.1$ rad exhibit features that are significantly different than those for other opening angles. This is because for $\alpha = \theta_s$ and $\phi_{k_i} = 0$, the incident projectile has approximately the same transverse momentum as the scattered projectile, resulting in only longitudinal momentum transferred to the target. Purely longitudinal momentum transfer then results in two peaks in the FDCS parallel (0°) and antiparallel (180°) to the beam direction.

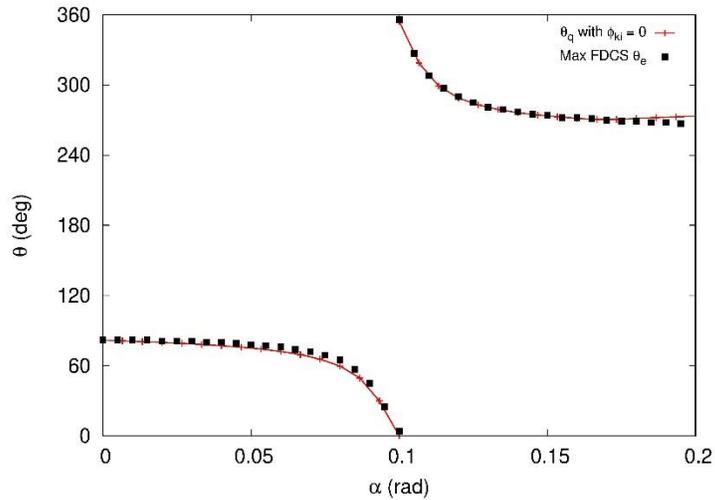

Figure 4 Momentum transfer angle (red line) as measured counterclockwise from the beam direction for incident momentum with azimuthal angle $\phi_{k_i} = 0$. Also shown is the ionized electron angle (black squares) corresponding to the maximum FDCS value. The incident projectile energy is 1 keV, scattering angle 100 mrad, ionized electron energy 5 eV, and ground state hydrogen target.

For the excited s-state targets, interference structures can be observed in the FDCS due to the multilobe structure of the target. While present for most values of $\alpha$, including $\alpha = 0$ non-vortex projectiles, the interference patterns are most noticeable for $\alpha$ near $\theta_s$. This is likely due to the spreading of the dominant lobe as the uncertainty in the momentum transfer increases, making the interference structures more visible. Unlike the s-state targets, the 2p target is not spherically symmetric and can serve as an analogue for diatomic molecule targets, where nuclear alignment effects are known to be important [31,32]. As seen from Fig. 2, the orientation of the target in the 2p state has a significant influence on the shape of the FDCS, although some of the features observed in the s-state FDCS persist.

For $\alpha < \theta_s$, the FDCS for 2pz orientation show a minimum along the classical momentum transfer direction with equal magnitude peaks on either side of this direction. This is because the 2pz plane wave transition matrix is minimized for the ejected electron momentum along the momentum transfer direction. For 2py orientation, the FDCS are zero at $\alpha = 0$ due to zero target wave function density in the scattering plane and the momentum transfer vector lying in the scattering plane. However, the use of a vortex projectile results in an out-of-plane component to the momentum transfer, which can then produce electrons ejected into the scattering plane, resulting in a single binary peak in the FDCS. A similar binary peak structure is seen in the 2px FDCS, which is enhanced for vortex projectiles.

One of the most interesting features of the 2p FDCS is the structure observed with $\alpha = \theta_s$. As for the s-states, the $\phi_{k_i} = 0$ FDCS is dominant and the primary momentum transfer direction is parallel to the beam direction. Therefore, any structures observed at ionized electron angles other than 0° or 180° must be due to the target electron's initial momentum or spatial distribution. For 2px orientation, two peaks in the FDCS are observed to either side of the classical beam direction, possibly caused by the target electron being offset to either side of the beam direction with zero density along the z-axis. This initial spatial distribution, combined with a momentum transfer vector along the beam direction, then results in an ionized electron distribution primarily located to either side of the beam direction. For 2py orientation, the ionized electrons are uniformly distributed due to the initial state electron distribution being symmetric about the scattering plane with zero initial state density in the scattering plane. For 2pz orientation, peaks are observed along and opposite the beam direction due to the momentum transfer along the beam direction and the electron density oriented along the beam direction. The strong dependence of the FDCS on 2p target orientation, along with the easily observable interference structures in the 2s and 3s FDCS provide preliminary evidence that the target orbital structure and orientation can be deduced from the FDCS using EV projectiles. This is a promising indicator that EV collisions may be used to characterize molecular structure.

## 4. Conclusion

FDCS for ionization of atomic hydrogen excited states by EV projectiles show signatures of atomic orbital structure and target orientation. Some of these features are not present in FDCS for non-vortex projectiles and others are more enhanced when EV projectiles are used. In particular, clear signatures of orientation effects were seen in the FDCS for 2p targets, and interference effects were observed for 2s and 3s targets resulting from their nodal structure. Analysis of the FDCS revealed that while the momentum of the incident projectile, and therefore the momentum transfer, is uncertain, the FDCS are dominated by an incident projectile with azimuthal angle $\phi_{k_i} = 0$. As the opening angle of the EV is varied, the uncertainty in the momentum transfer leads to a spreading of the ejected electron binary peak. The location of the binary peak was strongly correlated with the momentum transfer direction for an incident projectile with $\phi_{k_i} = 0$.

The results here demonstrate the potential feasibility of using ionization cross sections to infer target structure information, a requirement for some of the proposed applications of EV projectiles, such as characterization of chiral molecule enantiomers [13].  Our results also demonstrate a possible mechanism for controlling ionized electron emission angle by altering the EV opening angle.  The FDCS presented here provide valuable fundamental information for use in potential applications of EV projectile collisions and provide proof of principle that EV projectiles yield information not available or easily accessible by non-vortex projectiles.

## 5. Acknowledgements

We gratefully acknowledge the support of the NSF under Grant No. PHY-1912093.